\RequirePackage{fix-cm}
\documentclass[twocolumn,epjc3]{svjour3}  
\smartqed  
\RequirePackage{graphicx}
\newcommand{\be}{\begin{equation}}
\newcommand{\ee}{\end{equation}}
\newcommand{\bn}{\begin{eqnarray}}
\newcommand{\en}{\end{eqnarray}}
\newcommand{\bes}{\begin{subequations}}
\newcommand{\ees}{\end{subequations}}

\usepackage{epstopdf}
\usepackage{widetext}
\journalname{Eur. Phys. J. C}
\begin{document}

\title{The trace of the trace of the energy-momentum tensor-dependent Einstein's field equations}


\author{P.H.R.S. Moraes\thanksref{e1,addr1,addr2}
        }

\thankstext{e1}{e-mail: moraes.phrs@gmail.com}


\institute{ITA - Instituto Tecnol\'ogico de Aeron\'autica - Departamento de F\'isica, S\~ao Jos\'e dos Campos 12228-900, S\~ao Paulo, Brazil \label{addr1}
           \and
          UNINA - Università degli Studi di Napoli ``Federico II'' - Dipartimento di Fisica, Napoli I-80126, Italy \label{addr2}
}

\date{Received: date / Accepted: date}

\maketitle

\begin{abstract}

The $f(R,T)$ gravity field equations depend generically on both the Ricci scalar $R$ and trace of the energy-momentum tensor $T$. Within the assumption of perfect fluids, the theory carries an arbitrariness regarding the choice of the matter lagrangian density $\mathcal{L}$, not uniquely defined. Such an arbitrariness can be evaded by working with the trace of the theory field equations. From such an equation, one can obtain a form for $\mathcal{L}$,  which does not carry the arbitrariness. The obtained form for $\mathcal{L}$ shows that the $f(R,T)$ gravity is unimodular. A new version of the theory is, therefore, presented and forthcoming applications are expected. 

\end{abstract}

\keywords{$f(R,T)$ gravity \and matter lagrangian \and unimodular gravity}

\section{Introduction}\label{sec:int}

Most of the cosmological and astrophysical observational issues nowadays can be solved or evaded by assuming one of the following approaches: $i)$ a non-standard matter structure\footnote{The term ``non-standard matter structure'' may sound diffuse now but I hope to clarify it throughout the next paragraphs.} to be the content of the referred system; $ii)$ a non-standard underlying gravity theory. 

Take, as a first example, the accelerated expansion of the universe \cite{riess/1998,riess/2004,weinberg/2013}. In the standard model of cosmology, such a counter-intuitive phenomenon is described by {\it considering the existence of a  cosmological constant}, with negative pressure, filling the whole universe and being responsible for $\sim70\%$ of its composition \cite{planck_collaboration/2016,hinshaw/2013}. It is well-known that the cosmic acceleration may also be described with no need for invoking a cosmological constant, rather by {\it changing the underlying gravity theory}, that is, assuming an extension of General Relativity as the gravity theory for a cosmological model to be based on. This can be checked, for instance, in \cite{qiang/2005,sami/2016,lue/2004,piazza/2014,akarsu/2018}.

In the galactic scale it is known that in order to be in touch with observations of rotation curves, one has to {\it assume that most part of the galaxy is filled by dark matter} \cite{sofue/2012,kamada/2017,kent/1986,kent/1987}, a sort of matter that does not interact electromagnetically and therefore cannot be seen. On the other hand, {\it extensions of General Relativity} are able to describe galactic observations with no need for invoking dark matter \cite{o'brien/2018,o'brien/2012,mannheim/2012}. 

Another example comes from stellar astrophysics. Despite the well-known Chandrasekhar limit on white dwarf masses \cite{chandrasekhar/1931}, some super-Chandrasekhar white dwarfs have been observed \cite{tanaka/2010,silverman/2011,hachinger/2012,taubenberger/2013}. Those can be predicted by {\it changing the white dwarf standard structure,  assuming strong magnetic fields} inside them \cite{franzon/2015,chamel/2013,das/2013,das/2014,das/2014b,paret/2015,kawka/2017,wen/2014,terrero/2015} or by treating these objects in their hydrostatic equilibrium {\it within extended gravity theories}, in which upper Chandrasekhar limits are naturally obtained \cite{panah/2019,das/2015,das/2015b,kalita/2018}.

Some massive pulsars have also been detected \cite{antoniadis/2013,demorest/2010,linares/2018}. In order to particularly explain PSR J1614-2230, reported in \cite{demorest/2010}, in \cite{katayama/2012}, the tensor coupling of vector mesons to octet baryons was considered as well as the form factors at interaction vertices and a change in the internal (quark) structure of baryons in dense matter. Many other alternatives that take into account {\it changes in the superdense matter structure} inside neutron stars were considered to explain PSR J1614-2230 \cite{weissenborn/2012,weissenborn/2011,bednarek/2012,vidana/2011,bhar/2016}. It is important, here, to mention that within the simplest choices for equations of state of neutron stars, such as the polytropic one \cite{tooper/1964}, it is not possible to predict the existence of such massive pulsars in the context of General Relativity. On the other hand, some constraints on {\it scalar-tensor theories of gravity} from massive neutron stars can be appreciated in \cite{palenzuela/2016}.

This interface can be seen even in the physics of wormholes, which have not yet been detected, despite the efforts \cite{shaikh/2018,tsukamoto/2012,kuhfittig/2014,nandi/2017}. According to General Relativity, wormholes {\it must be filled by exotic (negative mass) matter} \cite{morris/1988,morris/1988b} while in {\it extended gravity} it is possible to obtain wormhole solutions with ordinary matter \cite{ms/2018,hohmann/2014}. 

Here I will choose the option of working with extensions of General Relativity. Particularly, the $f(R,T)$ gravity \cite{harko/2011} will be approached, for which $f(R,T)$ represents a function of the Ricci scalar $R$ and trace of the energy-momentum tensor $T$ to substitute $R$ in the Einstein-Hilbert gravitational action. 

The $f(R,T)$ gravity has been a common choice to underline the above issues, such as the accelerated expansion of the universe \cite{smsb/2018,pradhan/2018,kumar/2017,zaregonbadi/2016,mrc/2016,hossienkhani/2014,houndjo/2012}, dark matter \cite{zaregonbadi/2016b}, super-Chandrasekhar white dwarfs \cite{clmaomm/2017}, massive pulsars \cite{mam/2016} and wormholes \cite{sharif/2019,elizalde/2019,elizalde/2018,bhatti/2018,sms/2018,smsr/2018,ms/2017,mcl/2017,yousaf/2017,zubair/2016} (check also \cite{ms/2018}).

In this letter I will calculate the trace of the $f(R,T)$ gravity field equations. The resulting equation makes, from now on, unnecessary to choose a particular matter lagrangian density of perfect fluids within the formalism, which is an arbitrariness carried by the theory. It is revealed that the $f(R,T)$ gravity, in the absence of such an arbitrariness, is unimodular.

\section{The $f(R,T)$ gravity}\label{sec:frt}

The $f(R,T)$ gravitational theory total action reads as \cite{harko/2011}

\begin{equation}\label{frt1}
S=\int d^{4}x\sqrt{-g}\left[\frac{f(R,T)}{16\pi}+\mathcal{L}\right],
\end{equation}
for which the integration is made in the four-dimensional space-time\footnote{To get in touch with some extra-dimensional $f(R,T)$ models, check \cite{moraes/2015,mc/2016,cm/2016,pawar/2018,samanta/2017,sahoo/2016,ram/2013,reddy/2013,reddy/2012,gu/2017,dasunaidu/2018,sharif/2018,khan/2018,rao/2015,samanta/2013}.}, $g$ is the determinant of the metric $g_{\mu\nu}$, $\mathcal{L}$ is the matter lagrangian density and natural units are assumed.

From (\ref{frt1}), it is intuitive that the $f(R,T)$ gravity is allowed to have extra terms on both sides of Einstein's fields equations, namely, those coming from geometrical and material corrections. Here I will be concerned, particularly, with $f(R,T)$ models that contain only material correction terms, namely $f(R,T)=R+\mathcal{F}(T)$ models, with $\mathcal{F}(T)$ being a function of $T$ only. This is quite usual when one wishes to particularly investigate the role of extra material rather than geometrical terms in gravity. Anyhow, $f(R,T)$ models with correction terms on both sides of Einstein's field equations can be seen in \cite{smsb/2018,sms/2018,zubair/2016,cm/2016} and also \cite{zubair/2015,noureen/2015,amam/2016}. Furthermore, for an $f(R,T)$ model containing a strong coupling between geometry and matter, such as $f(R,T)=R+\alpha RT$, with constant $\alpha$, check \cite{ms/2017b}.

Now, by substituting $f(R,T)=R+\mathcal{F}(T)$ in (\ref{frt1}) and applying the variational principle, one obtains the following field equations 

\begin{equation}\label{frt2}
G_{\mu\nu}=8\pi T_{\mu\nu}+\frac{\mathcal{F}g_{\mu\nu}}{2}+\frac{d\mathcal{F}}{dT}(T_{\mu\nu}-\mathcal{L}g_{\mu\nu}).
\end{equation}

The extra terms\footnote{``Extra'' in comparison to General Relativity field equations.} in (\ref{frt2}) can be interpreted in different forms. Firstly, they can be related to fluid  imperfections. On this regard, note that even if one assumes that $T_{\mu\nu}$ is the energy-momentum tensor of a perfect fluid, the ``effective'' energy-momentum tensor of the theory will contain some extra terms that may be related to bulk viscosity, for instance. On this regard, some viscous fluid cosmological models can be seen in References \cite{meng/2009,mostaghel/2017,mohanty/1990}.

Secondly, they can be related to some quantum effects, such as the creation of particles in a quantum level. It is straightforward to see that by applying the Bianchi identities in (\ref{frt2}) yields $\nabla_\mu T^{\mu\nu}\neq0$, which, in a cosmological level, is interpreted as a mechanism of creation of particles throughout the universe evolution. Some cosmological models in a scenario with matter creation can be seen in \cite{quintin/2014,biswas/2017,maeda/1984}.

Thirdly, the extra terms appearing in Eq.(\ref{frt2}) may represent an extra fluid so that $f(R,T)$ gravity is effectively a two-fluid model, such as the models presented in \cite{evrard/1994,pradhan/2014,amirhashchi/2013}. 

Finally, the $f(R,T)$ models may represent, for a particular form of the function $\mathcal{F}(T)$ (check (\ref{frt2})), an effective cosmological constant model, with the difference that in $f(R,T)$ models the effective cosmological ``constant'' depends explicitly on $T$ rather than on time, such as the well-known decaying vacuum models \cite{freese/1987,rajantie/2017}. 

It is important to mention that departing from the non-$f(R,T)$ gravity models mentioned in References \cite{meng/2009}-\cite{rajantie/2017}, the $f(R,T)$ gravity is not based on any phenomenology. Rather the $T$-terms are inserted fundamentally on the gravitational action (\ref{frt1}). Nevertheless, the $T$-dependence of the theory may describe some fluid imperfections \cite{samanta/2017b}, quantum effects \cite{lcmm/2019,xu/2016}, two-fluid models \cite{mcr/2018,shabani/2018} and a varying cosmological constant \cite{tiwari/2017,singh/2016}.

A relevant point to be remarked on Eq.(\ref{frt2}) is that it depends explicitly on the choice of the matter lagrangian density $\mathcal{L}$.  In this way, different choices of the matter lagrangian yield different field equations. This is not satisfactory with respect to the monistic view of modern physics, which requires a unique mathematical description of natural phenomena.

\section{The trace of the $f(R,T)$ gravity field equations}\label{sec:tfrt}

The trace of Eq.(\ref{frt2}) reads

\begin{equation}\label{tfrt1}
-R=2(4\pi T+\mathcal{F})+\frac{d\mathcal{F}}{dT}(T-4\mathcal{L}).
\end{equation}

Let me consider the case of a perfect fluid, vastly applied in the cosmological and astrophysical contexts. Perfect fluids are defined by the matter-energy density $\rho$, pressure $p$ and four-velocity $u^\mu$, which must satisfy $u_\mu u^\mu=1$ and $u^\mu\nabla_\nu u_\mu=0$. 

It is known that the matter lagrangian density of a perfect fluid is not uniquely defined \cite{harko/2011,harko/2014,avelino/2018}. It is quite usual to see choices such as $\mathcal{L}=\rho$ and $\mathcal{L}=-p$. It is worth mentioning that this degeneracy has no consequences in General Relativity since in this case both lagrangians lead to the same field equations.

This degeneracy makes some conclusions about $f(R,T)$ cosmology \cite{velten/2017} to be, at least, premature, since the  approach could be reconstructed from a different choice for $\mathcal{L}$.  

Equation (\ref{tfrt1}) may solve such a degeneracy. By isolating $\mathcal{L}$ yields

\begin{equation}\label{mldug1}
\mathcal{L}=\frac{1}{4}\left\{T+\frac{dT}{d\mathcal{F}}[R+2(4\pi T+\mathcal{F})]\right\}.
\end{equation}

By substituting (\ref{mldug1}) in (\ref{frt2}) yields

\begin{equation}\label{mldug2}
R_{\mu\nu}-\frac{Rg_{\mu\nu}}{4}=\left(8\pi+\frac{d\mathcal{F}}{dT}\right)T_{\mu\nu}-\left(2\pi T+\frac{1}{4}\frac{d\mathcal{F}}{d\ln T}\right)g_{\mu\nu},
\end{equation} 
in which the definition of the Einstein tensor, namely $G_{\mu\nu}=R_{\mu\nu}-Rg_{\mu\nu}/2$, was used, with $R_{\mu\nu}$ being the Ricci tensor.

The use of (\ref{mldug1}) not only evades the need for choosing a particular matter lagrangian density but also attains the unimodular gravity \cite{ellis/2011,weinberg/1989,garcia-aspeitia/2019,anderson/1971,einstein/1919}, as one can check Eq.(\ref{mldug2}). The unimodular gravity has been applied, for instance, to solve the cosmological constant problem \cite{ellis/2011,weinberg/1989}, since in this approach the cosmological constant originates from the traceless part of the Einstein field equations as an integration constant and not as an input parameter, and to treat the accelerated expansion of the universe \cite{garcia-aspeitia/2019}. Note that the {\it lhs} of (\ref{mldug2}) reads exactly as the traceless Einstein tensor and by taking $\mathcal{F}=0$ one recovers exactly the unimodular gravity field equations. 

Remarkably, for any $\mathcal{F}$, Eq.(\ref{mldug2}) keeps traceless, what can be seen by calculating its trace. This is an indication that the $f(R,T)$ gravity is fundamentally unimodular, since by eliminating the arbitrariness usually carried by the theory, the formalism becomes unimodular.

Naturally, Eq.(\ref{mldug2}) serves as the base for new cosmological models to be implemented further on.

\section{Discussion}\label{sec:d}

In the present letter, by using the trace of the $f(R,T)$ gravity field equations, I have eliminated the arbitrariness carried by the $f(R,T)$ gravity on the choice of the matter lagrangian of a perfect fluid. Such an arbitrariness is also seen, for instance, in the $f(R,\mathcal{L})$ gravity \cite{harko/2010}, and naturally, the approach here can be reconstructed in the latter theory. For some interesting $f(R,\mathcal{L})$ gravity applications, one can check  \cite{azevedo/2016,wu/2014,harko/2013,wang/2012}.

The arbitrariness on the choice of the matter lagrangian of perfect fluids has been discussed earlier in the literature. Although $\mathcal{L}=\rho$ and $\mathcal{L}=-p$ are the most common choices, even $\mathcal{L}=T$ has already been seen  \cite{avelino/2018}. 

It is important to mention that in non-minimal geometry-matter coupling gravity models, such as the $f(R,T)$ and $f(R,\mathcal{L})$ theories, the motion of test particles submitted to gravitational fields is non-geodesic and happens in the presence of an extra force which is perpendicular to the four-velocity \cite{harko/2011,harko/2014,harko/2010,barrientos/2014}. The extra force depends on the matter lagrangian density and can even vanish in the case $\mathcal{L}=-p$ for the dust non-relativistic matter case. In this way, being such an extra force a measurable quantity, which can be related, for instance,  to the dark matter effects in rotation curves of galaxies \cite{faraoni/2009}, it is hardly likely that it carries an arbitrariness.

T. Harko has shown in \cite{harko/2010b} that the matter lagrangian in non-minimal geometry-matter coupling theories can be uniquely determined by the nature of the geometry-matter coupling, through a different approach than the present one, by considering the newtonian limit of the particle action for a fluid obeying a barotropic equation of state. He has shown that the matter lagrangian can be expressed either in terms of the density or pressure and in both cases the physical interpretation of the system is equivalent. In this way, the presence of the extra force is independent of the specific form of the matter lagrangian density and, in fact, it never vanishes.  

In \cite{minazzoli/2012}, O. Minazzoli and T. Harko have argued that the works that consider $\mathcal{L}=-p$ for their on-shell perfect fluid lagrangian or any linear combinations of $\rho$ and $-p$ may be incorrect as long as they deal with theories where the matter lagrangian density enters directly into the field equations. They have proved that this arbitrariness is incompatible with the matter current conservation. It was argued that it is very unlikely that even in the case in which the matter current conservation is relaxed, the lagrangian would reduce precisely to $-p$ or any linear combinations of $\rho$ and $-p$. These discussions clearly strengthen the approach presented in the present letter.

Remarkably, the present approach not only evaded the issue of the matter lagrangian density arbitrariness carried by non-minimal geometry-matter coupling theories, but also revealed that in a non-arbitrary scenario, the $f(R,T)$ gravity is unimodular, regardless of the function $\mathcal{F}(T)$.

A lot of attention has been given to unimodular gravity in the recent literature. Besides \cite{ellis/2011,garcia-aspeitia/2019}, one can check \cite{eichorn/2015} for the unimodular $f(R)$ gravity, \cite{bamba/2017} for the unimodular $f(\mathcal{T})$ gravity, with $\mathcal{T}$ being the torsion scalar, and \cite{bufalo/2015,alvarez/2015,saltas/2014,percacci/2018} for the unimodular quantum gravity.

As it was mentioned before, the cosmological constant in the Einstein's field equations of General Relativity may explain the observed accelerated expansion of the universe \cite{riess/1998}-\cite{hinshaw/2013}. It represents the gravitational effects of the quantum vacuum and consequently suffers from a strong fine-tuning problem \cite{weinberg/1989}. If one instead assumes the unimodular version of General Relativity, the vacuum energy has no gravitational effects. This does not determine a unique value for the cosmological constant, however it solves the fine tuning problem \cite{ellis/2011}. In \cite{ellis/2011}, G.F.R. Ellis et al. have also shown that the unimodular gravity also work for astrophysics (not only cosmology), such as the stellar equilibrium and black hole studies. Therefore, the unimodular gravity rises as an interesting alternative to General Relativity, which demands further investigations. The unimodular $f(R,T)$ gravity here obtained is one of the fields of unimodular gravity that worth further applications.

A first application could naturally be the development of the Friedmann-Lem\^aitre-Robertson-Walker cosmological model from the inception of the Friedmann-Lem\^aitre-Robertson-Walker metric as well as the energy-momentum tensor of a perfect fluid in Equation (\ref{mldug2}). The investigation of the other issues mentioned in Introduction, such as galactic rotation curves and hydrostatic equilibrium configurations of white dwarfs and neutron stars are also very much encouraged in this new unimodular $f(R,T)$ gravity scenario, free of the undesirable matter density lagrangian arbitrariness.

\begin{acknowledgements}
I would like to thank São Paulo Research Foundation (FAPESP), grants 2015/08476-0 and 2018/20689-7, for financial support. I also thank FAPESP under the thematic project 2013/26258-4. 
\end{acknowledgements}


\begin{thebibliography}{}

\bibitem{riess/1998} A.G. Riess et al., Astron. J. {\bf 116} (1998) 1009.
\bibitem{riess/2004} A.G. Riess et al., Astrophys. J. {\bf 607} (2004) 665.
\bibitem{weinberg/2013} D.H. Weinberg et al., Phys. Rep. {\bf 530} (2013) 87.
\bibitem{planck_collaboration/2016} Planck Collaboration, Astron. Astrophys. {\bf 594} (2016) A13.
\bibitem{hinshaw/2013} G. Hinshaw et al., Astrophys. J. {\bf 208} (2013) 19.
\bibitem{qiang/2005} L.-E Qiang et al., Phys. Rev. D {\bf 71} (2005) 061501.
\bibitem{sami/2016} M. Sami and R. Myrzakulov, Int. J. Mod. Phys. D {\bf 25} (2016) 1630031.
\bibitem{lue/2004} A. Lue et al., Phys. Rev. D {\bf 69} (2004) 044005.
\bibitem{piazza/2014} F. Piazza et al., J. Cosm. Astrop. Phys. {\bf 05} (2014) 043.
\bibitem{akarsu/2018} \"{O}. Akarsu et al., Phys. Rev. D {\bf 97} (2018) 024011.
\bibitem{sofue/2012} Y. Sofue, Publ. Astron. Soc. Jap. {\bf 64} (2012) 75.
\bibitem{kamada/2017} A. Kamada et al., Phys. Rev. Lett. {\bf 119} (2017) 111102.
\bibitem{kent/1986} S.M. Kent, Astron. J. {\bf 91} (1986) 1301.
\bibitem{kent/1987} S.M. Kent, Astron. J. {\bf 93} (1987) 816.
\bibitem{o'brien/2018} J.G. O'Brien et al., Astrophys. J. {\bf 852} (2018) 6.
\bibitem{o'brien/2012} J.G. O'Brien and P.D. Mannheim, Month. Not. Roy. Astron. Soc. {\bf 421} (2012) 1273.
\bibitem{mannheim/2012} P.D. Mannheim and J.G. O'Brien, Phys. Rev. D {\bf 857} (2012) 124020.
\bibitem{chandrasekhar/1931} S. Chandrasekhar, Astrophys. J. {\bf 74} (1931) 81.
\bibitem{tanaka/2010} M. Tanaka et al., Astrophys. J. {\bf 714} (2010) 1209.
\bibitem{silverman/2011} J.M. Silverman et al., Month. Not. Roy. Astron. Soc. {\bf 410} (2011) 585.
\bibitem{hachinger/2012} S. Hachinger et al., Month. Not. Roy. Astron. Soc. {\bf 427} (2012) 2057.
\bibitem{taubenberger/2013} S. Taubenberger et al., Month. Not. Roy. Astron. Soc. {\bf 432} (2013) 3117.
\bibitem{franzon/2015} B. Franzon and S. Schramm, Phys. Rev. D {\bf 92} (2015) 083006.
\bibitem{chamel/2013} N. Chamel et al., Phys. Rev. D {\bf 88} (2013) 081301.
\bibitem{das/2013} U. Das et al., Astrophys. J. {\bf 767} (2013) L14.
\bibitem{das/2014} U. Das and B. Mukhopadhyay, J. Cosm. Astrop. Phys. {\bf 06} (2014) 050.
\bibitem{das/2014b} U. Das and B. Mukhopadhyay, Mod. Phys. Lett. A {\bf 29} (2014) 1450035.
\bibitem{paret/2015} D.M. Paret et al., Res. Astron. Astrophys. {\bf 15} (2015) 1735.
\bibitem{kawka/2017} A. Kawka et al., Month. Not. Roy. Astron. Soc. {\bf 466} (2017) 1127.
\bibitem{wen/2014} D.-H. Wen et al., Chin. Phys. B {\bf 23} (2014) 089501.
\bibitem{terrero/2015} D.A. Terrero et al., Astron. Nach. {\bf 336} (2015) 851.
\bibitem{panah/2019} B.E. Panah and H.L. Liu, Phys. Rev. D {\bf 99} (2019) 104074.
\bibitem{das/2015} U. Das and B. Mukhopadhyay, Int. J. Mod. Phys. D {\bf 24} (2015) 1544026.
\bibitem{das/2015b} U. Das and B. Mukhopadhyay, J. Cosm. Astrop. Phys. {\bf 05} (2015) 045.
\bibitem{kalita/2018} S. Kalita and B. Mukhopadhyay, J. Cosm. Astrop. Phys. {\bf 09} (2018) 007.
\bibitem{antoniadis/2013} J. Antoniadis et al., Science {\bf 340} (2013) 448.
\bibitem{demorest/2010} P.B. Demorest et al., Nature {\bf 467} (2010) 1081.
\bibitem{linares/2018} M. Linares et al., Astrophys. J. {\bf 859} (2018) 54.
\bibitem{katayama/2012} T. Katayama et al., Astrophys. J. Supp. {\bf 203} (2012) 22.
\bibitem{weissenborn/2012} S. Weissenborn et al., Nucl. Phys. A {\bf 881} (2012) 62.
\bibitem{weissenborn/2011} S. Weissenborn et al., Astrophys. J. Lett. {\bf 740} (2011) L14.
\bibitem{bednarek/2012} I. Bednarek et al., Astron. Astrophys. {\bf 543} (2012) A157.
\bibitem{vidana/2011} I. Vida\~na et al., Europhys. Lett. {\bf 94} (2011) 11002.
\bibitem{bhar/2016} P. Bhar et al., Eur. Phys. J. A {\bf 52} (2016) 312.
\bibitem{tooper/1964} R.F. Tooper, Astrophys. J. Lett. {\bf 140} (1964) 434.
\bibitem{palenzuela/2016} C. Palenzuela and S.L. Liebling, Phys. Rev. D {\bf 93} (2016) 044009.
\bibitem{shaikh/2018} R. Shaikh, Phys. Rev. D {\bf 98} (2018) 024044.
\bibitem{tsukamoto/2012} N. Tsukamoto et al., Phys. Rev. D {\bf 86} (2012) 104062.
\bibitem{kuhfittig/2014} P.K.F. Kuhfittig, Eur. Phys. J. C {\bf 74} (2014) 2818.
\bibitem{nandi/2017} K.K. Nandi et al., Phys. Rev. D {\bf 95} (2017) 104011.
\bibitem{morris/1988} M.S. Morris and K.S. Thorne, Amer. J. Phys. {\bf 56} (1988) 395.
\bibitem{morris/1988b} M.S. Morris et al., Phys. Rev. Lett. {\bf 61} (1988) 1446.
\bibitem{ms/2018} P.H.R.S. Moraes and P.K. Sahoo, Phys. Rev. D {\bf 97} (2018) 024007.
\bibitem{hohmann/2014} M. Hohmann, Phys. Rev. D {\bf 89} (2014) 087503.
\bibitem{harko/2011} T. Harko et al., Phys. Rev. D {\bf 84} (2011) 024020.
\bibitem{smsb/2018} P.K. Sahoo, P.H.R.S. Moraes, P. Sahoo and B.K. Bishi, Eur. Phys. J. C {\bf 78} (2018) 736.
\bibitem{pradhan/2018} A. Pradhan and R. Jaiswal, Int. J. Geom. Meth. Mod. Phys. {\bf 15} (2018) 1850076-320.
\bibitem{kumar/2017} R.S. Kumar and B. Satyannarayana, Ind. J. Phys. {\bf 91} (2017) 1293.
\bibitem{zaregonbadi/2016} R. Zaregonbadi and M. Farhoudi, Gen. Rel. Grav. {\bf 48} (2016) 142.
\bibitem{mrc/2016} P.H.R.S. Moraes, G. Ribeiro and R.A.C. Correa, Astrophys. Spa. Sci. {\bf 361} (2016) 227.
\bibitem{hossienkhani/2014} H. Hossienkhani et al., Astrophys. Spa. Sci. {\bf 353} (2014) 311.
\bibitem{houndjo/2012} M.J.S. Houndjo, Int. J. Mod. Phys. D {\bf 21} (2012) 1250003-1.
\bibitem{zaregonbadi/2016b} R. Zaregonbadi et al., Phys. Rev. D {\bf 94} (2016) 084052.
\bibitem{clmaomm/2017} G.A. Carvalho, R.V. Lobato, P.H.R.S. Moraes, J.D.V. Arba\~nil, E. Otoniel, R.M. Marinho and M. Malheiro, Eur. Phys. J. C {\bf 77} (2017) 871.
\bibitem{mam/2016} P.H.R.S. Moraes, J.D.V. Arba\~nil and M. Malheiro, J. Cosm. Astrop. Phys. {\bf 06} (2016) 005.
\bibitem{sharif/2019} M. Sharif and I. Nawazish, Ann. Phys. {\bf 400} (2019) 37.
\bibitem{elizalde/2019} E. Elizalde and M. Khurshudyan, Phys. Rev. D {\bf 99} (2019) 024051.
\bibitem{elizalde/2018} E. Elizalde and M. Khurshudyan, Phys. Rev. D {\bf 98} (2018) 123525.
\bibitem{bhatti/2018} M.Z. Bhatti et al., J. Astrophys. Astron. {\bf 39} (2018) 69.
\bibitem{sms/2018} P.K. Sahoo, P.H.R.S. Moraes and P. Sahoo, Eur. Phys. J. C {\bf 78} (2018) 46.
\bibitem{smsr/2018} P.K. Sahoo, P.H.R.S. Moraes, P. Sahoo and G. Ribeiro, Int. J. Mod. Phys. D {\bf 27} (2018) 1950004.
\bibitem{ms/2017} P.H.R.S. Moraes and P.K. Sahoo, Phys. Rev. D {\bf 96} (2017) 044038.
\bibitem{mcl/2017} P.H.R.S. Moraes, R.A.C. Correa and R.V. Lobato, J. Cosm. Astrop. Phys. {\bf 07} (2017) 029.
\bibitem{yousaf/2017} Z. Yousaf et al., Eur. Phys. J. Plus {\bf 132} (2017) 268.
\bibitem{zubair/2016} M. Zubair et al., Eur. Phys. J. C {\bf 76} (2016) 444.
\bibitem{moraes/2015} P.H.R.S. Moraes, Eur. Phys. J. C {\bf 75} (2015) 168.
\bibitem{mc/2016} P.H.R.S. Moraes and R.A.C. Correa, Astrophys. Spa. Sci. {\bf 361} (2016) 91.
\bibitem{cm/2016} R.A.C. Correa and P.H.R.S. Moraes, Eur. Phys. J. C {\bf 76} (2016) 100.
\bibitem{pawar/2018} D.D. Pawar et al., New Astron. {\bf 65} (2018) 1.
\bibitem{samanta/2017} G.C. Samanta et al, Zeits. Naturf. {\bf 72} (2017) 365.
\bibitem{sahoo/2016} P.K. Sahoo et al., Ind. J. Phys. {\bf 90} (2016) 485.
\bibitem{ram/2013} S. Ram and Priyanka, Astrophys. Spa. Sci. {\bf 347} (2013) 389.
\bibitem{reddy/2013} D.R.K. Reddy et al., Astrophys. Spa. Sci. {\bf 346} (2013) 261.
\bibitem{reddy/2012} D.R.K. Reddy et al., Int. J. Theor. Phys. {\bf 51} (2012) 3222.
\bibitem{gu/2017} B.-M. Gu et al., Eur. Phys. J. C {\bf 77} (2017) 115.
\bibitem{dasunaidu/2018} K. Dasunaidu et al., Astrophys. Spa. Sci. {\bf 363} (2018) 158.
\bibitem{sharif/2018} M. Sharif and A. Anwar, Astrophys. Spa. Sci. {\bf 363} (2018) 123.
\bibitem{khan/2018} S. Khan et al., Mod. Phys. Lett. A {\bf 33} (2018) 1850065.
\bibitem{rao/2015} V.U.M. Rao and D.C.P. Rao, Astrophys. Spa. Sci. {\bf 357} (2015) 65.
\bibitem{samanta/2013} G.C. Samanta and S.N. Dhal, Int. J. Theor. Phys. {\bf 52} (2013) 1334.
\bibitem{zubair/2015} M. Zubair and I. Noureen, Eur. Phys. J. C {\bf 75} (2015) 265.
\bibitem{noureen/2015} I. Noureen et al., Eur. Phys. J. C {\bf 75} (2015) 323.
\bibitem{amam/2016} M.E.S. Alves, P.H.R.S. Moraes, J.C.N. de Araujo and M. Malheiro, Phys. Rev. D {\bf 94} (2016) 024032.
\bibitem{ms/2017b} P.H.R.S. Moraes and P.K. Sahoo, Eur. Phys. J. C {\bf 77} (2017) 480.
\bibitem{meng/2009} X.-H. Meng and X. Dou, Comm. Theor. Phys. {\bf 52} (2009) 377.
\bibitem{mostaghel/2017} B. Mostaghel et al., Eur. Phys. J. C {\bf 77} (2017) 541.
\bibitem{mohanty/1990} G. Mohanty and B.D. Pradhan, Astrophys. Spa. Sci. {\bf 165} (1990) 163.
\bibitem{quintin/2014} J. Quintin et al., Phys. Rev. D {\bf 90} (2014) 063507.
\bibitem{biswas/2017} S. Kr. Biswas et al., Phys. Rev. D {\bf 95} (2017) 103009.
\bibitem{maeda/1984} K.-I. Maeda, Phys. Rev. D {\bf 30} (1984) 2482.
\bibitem{evrard/1994} A.E. Evrard et al., Astrophys. J. {\bf 422} (1994) 11.
\bibitem{pradhan/2014} A. Pradhan, Ind. J. Phys. {\bf 88} (2014) 215.
\bibitem{amirhashchi/2013} H. Amirhashchi et al., Int. J. Theor. Phys. {\bf 52} (2013) 2735.
\bibitem{freese/1987} K. Freese et al., Nucl. Phys. B {\bf 287} (1987) 797.
\bibitem{rajantie/2017} A. Rajantie and S. Stopyra, Phys. Rev. D {\bf 95} (2017) 025008.
\bibitem{samanta/2017b} G.C. Samanta and R. Myrzakulov, Chin. J. Phys. {\bf 55} (2017) 1044.
\bibitem{lcmm/2019} R.V. Lobato, G.A. Carvalho, A.G. Martins and P.H.R.S. Moraes, Eur. Phys. J. Plus {\bf 134} (2019) 132.
\bibitem{xu/2016} M.-X. Xu et al., Eur. Phys. J. C {\bf 76} (2016) 449.
\bibitem{mcr/2018} P.H.R.S. Moraes, R.A.C. Correa and G. Ribeiro, Eur. Phys. J. C {\bf 78} (2018) 192.
\bibitem{shabani/2018} H. Shabani and A.H. Ziaie, Int. J. Mod. Phys. A {\bf 33} (2018) 1850050.
\bibitem{tiwari/2017} R.K. Tiwari et al., Astrophys. Spa. Sci. {\bf 362} (2017) 143.
\bibitem{singh/2016} G.P. Singh et al., Chin. J. Phys. {\bf 54} (2016) 244.
\bibitem{harko/2014} T. Harko and F.S.N. Lobo, Galaxies {\bf 2} (2014) 410.
\bibitem{avelino/2018} P.P. Avelino and R.P.L. Azevedo, Phys. Rev. D {\bf 97} (2018) 064018.
\bibitem{velten/2017} H. Velten and T.R.P. Caram\^es, Phys. Rev. D {\bf 95} (2017) 123536.
\bibitem{ellis/2011} G.F.R. Ellis et al., Class. Quant. Grav. {\bf 28} (2011) 225007.
\bibitem{weinberg/1989} S. Weinberg, Rev. Mod. Phys. {\bf 61} (1989) 1.
\bibitem{garcia-aspeitia/2019} M.A. Garc\'ia-Aspeitia et al., Phys. Rev. D {\bf 99} (2019) 123525.
\bibitem{anderson/1971} J.L. Anderson and D. Finkelstein, Am. J. Phys. {\bf 39} (1971) 901.
\bibitem{einstein/1919} A. Einstein, Sitzungsber. Preuss. Akad. Wiss. Berlin (Math. Phys.) (1919) 349.
\bibitem{harko/2010} T. Harko and F.S.N. Lobo, Eur. Phys. J. C {\bf 70} (2010) 373.
\bibitem{azevedo/2016} R.P.L. Azevedo and J. P\'aramos, Phys. Rev. D {\bf 94} (2016) 064036.
\bibitem{wu/2014} Y.-B. Wu et al., Mod. Phys. Lett. A {\bf 29} (2014) 1450089.
\bibitem{harko/2013} T. Harko et al., Phys. Rev. D {\bf 87} (2013) 047501.
\bibitem{wang/2012} J. Wang and K. Liao, Class. Quant. Grav. {\bf 29} (2012) 215016.
\bibitem{barrientos/2014} J. Barrientos O. and G.F. Rubilar, Phys. Rev. D {\bf 90} (2014) 028501.
\bibitem{faraoni/2009} V. Faraoni, Phys. Rev. D {\bf 80} (2009) 124040.
\bibitem{harko/2010b} T. Harko, Phys. Rev. D {\bf 81} (2010) 044021.
\bibitem{minazzoli/2012} O. Minazzoli and T. Harko, Phys. Rev. D {\bf 86} (2012) 087502.
\bibitem{eichorn/2015} A. Eichorn, J. High Ener. Phys. {\bf 2015} (2015) 96.
\bibitem{bamba/2017} K. Bamba et al., Mod. Phys. Lett. A {\bf 32} (2017) 1750114.
\bibitem{bufalo/2015} R. Bufalo et al., Eur. Phys. J. C {\bf 75} (2015) 477.
\bibitem{alvarez/2015} E. \'Alvarez et al., J. High Ener. Phys. {\bf 2015} (2015) 78.
\bibitem{saltas/2014} I.D. Saltas, Phys. Rev. D {\bf 90} (2014) 124052.
\bibitem{percacci/2018} R. Percacci, Found. Phys. {\bf 48} (2018) 1364.

\end{thebibliography}
\end{document}